\documentclass[amsmath,amssymb,12pt,nofootinbib,onecolumn]{revtex4}

\usepackage[dvipdfmx]{graphicx}
\usepackage{dcolumn}
\usepackage{bm}
\usepackage{color}
\usepackage{enumitem}
  \usepackage{tabularx}
   \newcolumntype{C}{>{\centering\arraybackslash}X}
   \newcolumntype{L}{>{\raggedright\arraybackslash}X}
   \newcolumntype{R}{>{\raggedleft\arraybackslash}X}
\setlistdepth{10}

\usepackage[normalem]{ulem}

\newcommand{\ii}{\mathrm{i}}

\newcommand{\del}{\partial}
\newcommand{\ee}{{\rm e}}

\definecolor{DarkBlue}{rgb}{0,0,0.7} 

\definecolor{DarkRed}{rgb}{0.65,0,0}

\begin{document}
\baselineskip5.5mm

{\baselineskip0pt
\small
\leftline{\baselineskip16pt\sl\vbox to0pt{
                             \vss}}
\rightline{\baselineskip16pt\rm\vbox to20pt{
\vspace{-1.5cm}
\vss}}
}

\vskip-1.cm

\author{Chul-Moon~Yoo}\email{yoo.chulmoon.k6@f.mail.nagoya-u.ac.jp}
\affiliation{
Division of Particle and Astrophysical Science,
Graduate School of Science, \\Nagoya University, 
Nagoya 464-8602, Japan
}

\title{Primordial black hole formation from a nonspherical \\density profile with a misaligned deformation tensor}

 \vspace{1cm}


\begin{abstract}
\baselineskip5.5mm 
We perform the numerical simulation of primordial black hole formation from 
a nonspherical profile of the initial curvature perturbation $\zeta$. 
We consider the background expanding universe filled with 
the perfect fluid with the linear equation of state $p=w\rho$ ($w=1/3$ or $1/5$), 
where $p$ and $\rho$ are the pressure and the energy density, respectively.  
The initial condition is set in a way such that 
the principal directions of the second derivatives of $\zeta$ and $\triangle \zeta$ 
at the central peak are misaligned, where $\triangle$ is the Laplacian. 
In this setting, since the linearized density is proportional to $\triangle \zeta$,  
the inertia tensor and deformation tensor $\del_i\del_j \zeta$ are misaligned.  
Thus tidal torque may act and the spin of a resultant primordial black hole 
would be non-zero in general, 
although it is estimated to be very small from previous perturbative analyses.
As a result, 
we do not find a finite value of the spin within our numerical precision, 
giving support for the negligibly small value of the 
black hole spin for $1/5\lesssim w \lesssim 1/3$. 
More specifically, our results suggest that the dimensionless PBH spin $s$ is typically so small that 
$s\ll0.1$ for $w\gtrsim0.2$. 
\end{abstract}


\maketitle
\thispagestyle{empty}
\pagebreak

\section{Introduction}
The possibility of black hole formation in the early universe has been 
proposed in Refs.~\cite{1967SvA....10..602Z,Hawking:1971ei,Carr:1974nx}
more than half a century ago. 
Since then, it had been a fascinating but relatively minor possible scenario until the first detection of the gravitational waves from a binary black hole~\cite{Abbott:2016blz}. 
However, after the possibility for the binary to originate from primordial black holes (PBHs) is pointed out~\cite{Sasaki:2016jop}, 
people started to realize the utility of PBHs for 
many kinds of subjects in cosmology and astrophysics. 

PBHs are the remnants 
of primordial inhomogeneity and there is no doubt that black holes 
can be formed if sufficiently over-dense regions exist in the early universe. 
They may play the role of dark matter~\cite{Carr:2009jm,Carr:2020gox,Carr:2020xqk} or could be the source of micro-lensing events~\cite{Niikura:2019kqi} 
and/or black hole binaries observed by the gravitational waves~\cite{Bird:2016dcv,Sasaki:2016jop,Clesse:2017bsw,Sasaki:2018dmp}. 
In addition, they may be the seeds of supermassive black holes and cosmic structures~\cite{Kawasaki:2012kn,Kohri:2014lza,Nakama:2016kfq,Carr:2018rid,Serpico:2020ehh,Unal:2020mts,Kohri:2022wzp}. 
Another attractive observable associated with large primordial perturbations is 
the gravitational waves induced by the primordial curvature perturbations~\cite{Ananda:2006af,Baumann:2007zm,Saito:2008jc,Saito:2009jt,Assadullahi:2009nf,Bugaev:2009zh,Bugaev:2010bb,Espinosa:2018eve,Kohri:2018awv,Domenech:2021ztg}. 
Since the induced gravitational waves and PBHs share 
the primordial curvature perturbations as those origins, 
combination of those observations are attractive tools to probe the early universe. 

In this paper, we focus on the PBH formation 
in a nonspherical setting and possible generation of the black hole spin. 
Although 
spherically symmetric dynamics for PBH formation has been mainly considered~\cite{Niemeyer:1997mt,Niemeyer:1999ak,Shibata:1999zs,2002CQGra..19.3687H,Musco:2004ak,Harada:2013epa,Nakama:2013ica,Nakama:2014fra,Harada:2015yda,Musco:2018rwt,Escriva:2019nsa,Escriva:2019phb,Escriva:2020tak,Escriva:2021aeh,Musco:2021sva,Escriva:2022bwe,Franciolini:2022tfm,Papanikolaou:2022cvo,Escriva:2023qnq,Uehara:2024yyp} for many years, 
we can also find several works in which nonspherical PBH formation is considered~\cite{Khlopov:1980mg,Harada:2016mhb,Chiba:2017rvs,Harada:2017fjm,DeLuca:2019buf,Harada:2020pzb,Yoo:2020lmg,deJong:2021bbo,Saito:2023fpt,deJong:2023gsx}. 
Some of them~\cite{Khlopov:1980mg,Harada:2016mhb,Harada:2017fjm} focus on the PBH formation in a matter-dominated universe, in which 
the non-spherical collapse would be essential for understanding the criterion of the PBH formation. 
In the case of a radiation-dominated universe~\cite{Chiba:2017rvs,DeLuca:2019buf,Harada:2020pzb,Yoo:2020lmg}, the pressure gradient is the main obstacle to PBH formation and a large amplitude of the curvature perturbation is required. 
Then, according to the peak theory~\cite{1986ApJ...304...15B}, the system approaches 
a spherically symmetric configuration in the high peak limit, 
so that the deviation from the spherical configuration becomes relatively ineffective. 
The situation would not change much for the cases of somewhat softer equations of states~\cite{Saito:2023fpt}. 

The previous works~\cite{DeLuca:2019buf,Harada:2020pzb,Saito:2023fpt} suggest that, for the case of a perfect fluid with non-negligible pressure, the PBH spin is negligibly small immediately after the formation. 
Nevertheless, the PBH spin has been estimated based on the perturbative analyses 
and non-linear simulation with a standard initial setting has not been done yet (see Refs.~\cite{deJong:2021bbo,deJong:2023gsx} for simulations with massive and massless scalar fields).
The purpose of this work is to check the validity of the results obtained in Refs.~\cite{DeLuca:2019buf,Harada:2020pzb,Saito:2023fpt} through fully non-linear numerical simulations. 
Therefore, following the setting in Refs.~\cite{DeLuca:2019buf,Harada:2020pzb,Saito:2023fpt}, we consider only one field variable, which corresponds to the growing mode solution of the curvature perturbation in the long-wavelength limit, characterizing the initial condition. 
Then we suppose the tidal torque as the mechanism to generate angular momentum transfer. 
That is, we do not explicitly introduce the initial angular momentum in an ad-hoc manner, 
and check the efficiency of the tidal torque in the PBH formation.

Throughout this paper, we use the geometrized units in which both 
the speed of light and Newton's gravitational constant are set to unity, $c=G=1$.

\section{Basic setups}
\label{sec:setup}

First, let us introduce the characteristic comoving length scale $1/k$ for 
the initial perturbation. 
Since we are interested in the perturbations which are initially
super-horizon scale, we assume $\epsilon:=k/(a_\ii H_\ii)\ll 1$ with $a_\ii$ and $H_\ii$ being 
the scale factor and the Hubble expansion rate at the initial time, respectively. 
For simplicity, hereafter we assume that the matter field is given by the perfect fluid with the linear equation of states $p=w\rho$, where $p$ and $\rho$ are the pressure and the energy density of the fluid. 
In this paper, we consider the two specific values $1/3$ and $1/5$ for $w$.  
Growing mode solutions for super-horizon scale perturbations can be obtained 
by performing the gradient expansion~\cite{Shibata:1999zs,Harada:2015yda} and it turns out that,
at the leading order,  
the solution can be characterized by the conformal factor of the spatial metric given as an arbitrary function of spatial coordinates $x^i$. 
Following a convention, we write the spatial metric of the initial condition as 
$\ee^{-2\zeta(\bm x)}a_\ii^2\eta_{ij}$, where $\eta_{ij}$ is the reference flat metric and 
$\zeta$ is an arbitrary function of the spatial coordinates $\bm x$. 
We simply call $\zeta$ the curvature perturbation hereafter. 

In practice, setting the functional form of $\zeta$, we calculate all geometrical variables 
following Ref.~\cite{Harada:2015yda} with the uniform Hubble (constant-mean-curvature) slicing and 
the normal coordinates (vanishing shift vectors). 
Then the fluid variables are calculated by using the exact form of the Hamiltonian and momentum constraint equations. 
That is, the fluid configuration is set in a way such that the constraint equations are satisfied within the machine's precision. 
The configuration is also consistent with the long-wavelength approximation 
as long as the typical length scale $a_\ii/k$ of the perturbation is sufficiently larger than the Hubble length $1/H_\ii$. 

\begin{figure}[htbp]
  \begin{center}
  \includegraphics[scale=0.4]{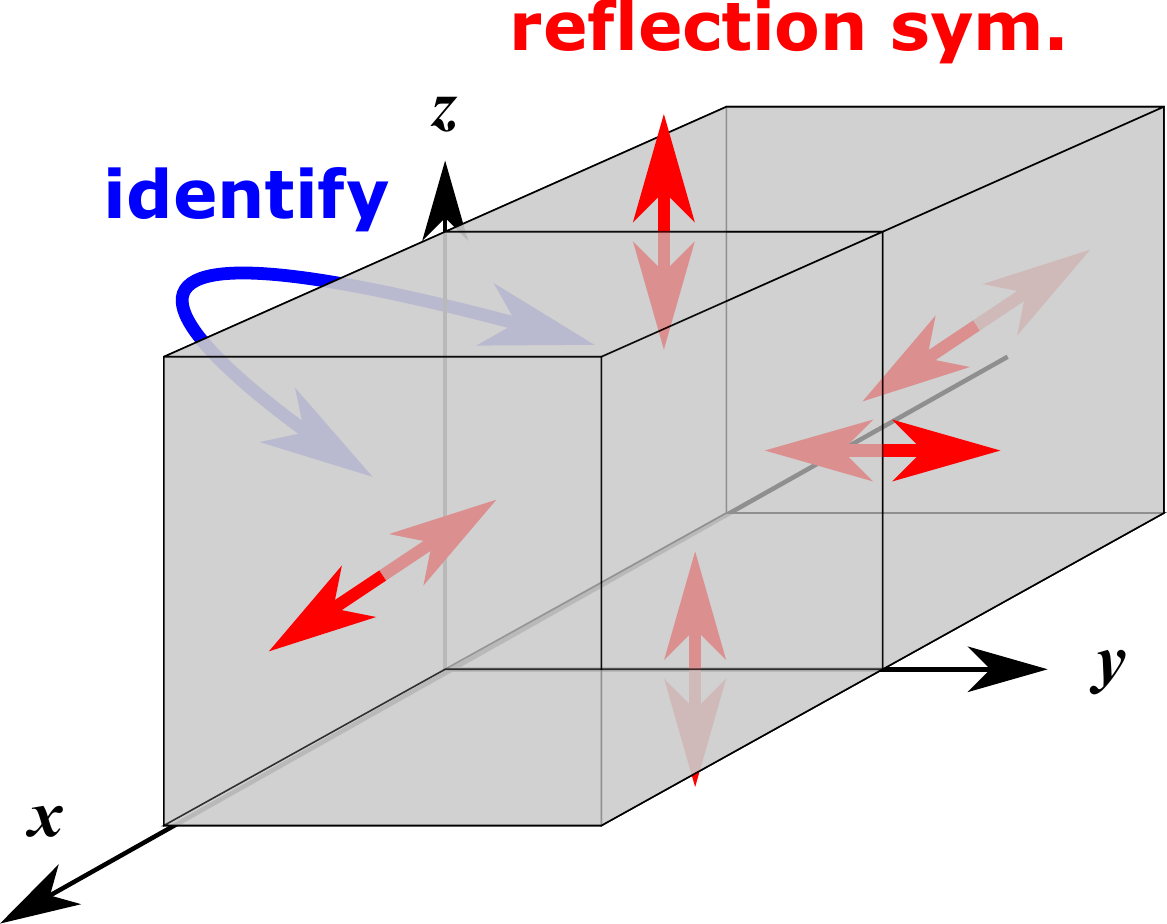}
  \caption{The numerical region and boundary conditions. 
  }
  \label{fig:boundaries}
  \end{center}
  \end{figure}

For the numerical simulation, we only consider the numerical domain given by 
$-L\leq X\leq L$, $0\leq Y\leq L$ and $0\leq Z\leq L$, where $X$, $Y$ and $Z$ are the 
reference Cartesian coordinates. 
We note that the Cartesian coordinates can be different 
from the reference spatial coordinates $x^i$ in general. 
Indeed, we will introduce scale-up coordinates as $x^i$ later. 
We identify the boundary surface $(0<X\leq L,Y=0,0\leq Z\leq L)$ as 
$(-L\leq X< 0,Y=0,0\leq Z\leq L)$ with 
the $\pi$ rotation around $Z$ axis (see Fig.~\ref{fig:boundaries}). 
The reflection symmetries are assumed for other boundary surfaces. 
The numerical region and the boundary conditions are summarized in Fig.~\ref{fig:boundaries}. 

In this paper, we use the following specific functional form of $\zeta$:
\begin{eqnarray}
  \zeta&=&-\mu\Bigl[1+\frac{1}{2}\left(k_1^2(X+Y)^2/2+k_2^2(X-Y)^2/2+k_3^2Z^2\right)
  \cr&+&\frac{1}{4}\left(k_1^2(X+Y)^2/2+k_2^2(X-Y)^2/2+k_3^2Z^2\right)^2\cr&+&\frac{1}{280}k^2R^2\left(9\kappa_1^2-\kappa_2^2-\kappa_3^2\right)X^2+(\kappa_1^2-9\kappa_2^2+\kappa_3^2)Y^2+(\kappa_1^2+\kappa_2^2-9\kappa_3^2)Z^2\Bigr]^{-1}\cr&&\exp\left[-\frac{1}{2880}k^6R^6\right], 
\end{eqnarray}
where 
\begin{eqnarray}
  R^2&=&X^2+Y^2+Z^2,\\
  k_1^2&=&\frac{1}{3}\left(\hat\xi_1+3\hat\xi_2+\hat\xi_3\right),\\
k_2^2&=&\frac{1}{3}\left(\hat\xi_1-3\hat\xi_2+\hat\xi_3\right),\\
k_3^2&=&\frac{1}{3}\left(\hat\xi_1-2\hat\xi_3\right),\\
\kappa_1^2&=&\frac{1}{3}\left(\widetilde \xi_1+3\widetilde \xi_2+\widetilde \xi_3\right),\\
\kappa_2^2&=&\frac{1}{3}\left(\widetilde \xi_1-3\widetilde \xi_2+\widetilde \xi_3\right),\\
\kappa_3^2&=&\frac{1}{3}\left(\widetilde \xi_1-2\widetilde \xi_3\right) 
\end{eqnarray}
with
\begin{eqnarray}
  &&k^2=\sum_{i=1,2,3}k_i^2=\hat\xi_1=\sum_{i=1,2,3}\kappa_i^2=\widetilde \xi_1=100/L^2,\cr
  &&\hat\xi_2=10/L^2,~\widetilde \xi_2=15/L^2,~\hat\xi_3=\widetilde \xi_3=0. 
\end{eqnarray}
There are some reasons for using this specific form of the curvature perturbation $\zeta$. 

First, by expanding $\zeta$ around the origin, we can find 
\begin{eqnarray}
  \frac{\zeta}{\mu}&\simeq&-1+\frac{1}{2}\left(k_1^2(X+Y)^2/2+k_2^2(X-Y)^2/2+k_3^2Z^2\right)+\mathcal O(R^4), \\
  \frac{\triangle \zeta}{\mu k^2}&\simeq&1-\frac{1}{2}\left(\kappa_1^2X^2+\kappa_2^2Y^2+\kappa_3^2Z^2\right)+\mathcal O(R^4), 
\end{eqnarray}
where $\triangle$ is the Laplacian for the reference flat metric. 
Since the linear density perturbation $\delta$ is proportional to $\triangle \zeta$, 
the principal directions of the deformation tensor $\del_i\del_j\zeta$ and the inertia tensor 
are misaligned with each other with $\pi/4$. 
Because of this misalignment, we may expect the tidal torque acts and 
the gravitational collapse is accompanied by the rotation~\cite{DeLuca:2019buf,Harada:2020pzb}. 
The last factor $\exp\left[-\frac{1}{2880}k^6R^6\right]$ is multiplied to realize $\zeta\simeq0$ around 
the outer boundary. In practice, we also introduce a window function as in Ref.~\cite{Yoo:2018pda} 
to regularize $\zeta$ in the vicinity of the outer boundary (see Eq.~(24) in Ref.~\cite{Yoo:2018pda} 
for the specific form of the window function). 

The eigenvalues of the inertia and deformation tensors $k_i$ and $\kappa_i$ 
are characterized by $\hat\xi_i$ and $\widetilde \xi_i$. 
The reason for using this expression is in the probability distribution 
based on the peak theory for those parameters~\cite{1986ApJ...304...15B}. 
Let us introduce the $n$-th order gradient moment $\sigma_n$~\cite{1986ApJ...304...15B}. 
According to the peak theory, the values of 
$-\zeta|_{\rm peak}/\sigma_0$, $\triangle\zeta|_{\rm peak}/\sigma_2$ and $-\triangle \triangle \zeta|_{\rm peak}/\sigma_4$
are significantly correlated with each other (see Eq.~(A7) in Ref.~\cite{Yoo:2020dkz} for the probability distribution function), where $~|_{\rm peak}$ indicates the value at the peak.  
From these correlations, we expect that the orders of magnitude for those variables are similar to each other. 
Here we simply assumed 
$-\zeta k^2\kappa^2=\triangle \zeta \kappa^2=-\triangle \triangle \zeta=\mu k^2\kappa^2$ at the peak with $k^2=\hat\xi_1=\kappa^2=\widetilde \xi_1$. 
The probability distribution of $\xi_2:=\hat\xi_2/\sigma_2$ and 
$\xi_3:=\hat \xi_3/\sigma_2$ are given by 
\begin{equation}
  P(\xi_2,\xi_3)=\frac{5^{5/2}3^2}{\sqrt{2\pi}}\xi_2(\xi_2^2-\xi_3^2)\exp\left[-\frac{5}{2}(3\xi_2^2+\xi_3^2)\right]
\end{equation}
without correlation with $-\zeta|_{\rm peak}/\sigma_0$ and  $\triangle\zeta|_{\rm peak}/\sigma_2$ (see, e.g., Eqs.(2-6) in Ref.~\cite{Yoo:2020lmg} with $\lambda_2\leftrightarrow\lambda_3$). 
The parameter region of $\xi_2$ and $\xi_3$ are restricted to 
$-\xi_2\leq\xi_3\leq-\xi_2$ and $0\leq\xi_2$. 
Then we choose the most probable value for $\xi_3$, namely, $\xi_3=0$. We can make a similar argument for $\widetilde \xi_3$ 
assigning $\triangle \zeta$ to the role of $\zeta$ in the above discussion. Thus we set $\hat \xi_3=\widetilde \xi_3=0$. 
Then the probability distribution for $\xi_2$ takes the maximum value at $\xi_2=1/\sqrt{5}\simeq 0.45$. 
Therefore we can roughly estimate the typical value of $\hat \xi_2$ as $\hat \xi_2=\xi_2\sigma_2\sim k^2\sigma/\sqrt{5}=100\sigma/(\sqrt{5}L^2)$ with $\sigma^2$ being the typical amplitude of the curvature power spectrum.  
Since, for PBH formation scenario, 
the value of $\sigma$ is typically given by $\sigma< 0.1$, 
we may evaluate the value of $\hat \xi_2$ as $\hat \xi_2\lesssim 5/L^2$. 
Therefore the values of $\hat\xi_2=10/L^2$ and $\widetilde\xi_2=15/L^2$ are unexpectedly large. 
These large values of $\hat\xi_2$ and $\widetilde \xi_2$ together with the finely tuned misalignment angle $\pi/4$ are assumed to make the tidal torque more effective. 
That is, the setting is optimized for the generation of the PBH spin, and the value of the PBH spin 
is expected to be larger than the typical value. 

\section{time evolution}
\label{sec:time_evo}

Our simulation code is based on the COSMOS code developed in Refs.~\cite{Yoo:2013yea,Okawa:2014nda}. 
For the simulation, we follow the numerical schemes adopted in Ref.~\cite{Yoo:2020lmg}, newly implementing a mesh-refinement procedure in the central region. 
The 4th-order Runge-Kutta method with the BSSN (Baumgarte-Shapiro-Shibata-Nakamura) formalism is 
used for solving the Einstein equations~\cite{Shibata:1995we,Baumgarte:1998te}. 
For the spatial coordinates, we employ the scale-up coordinates introduced in Ref.~\cite{Yoo:2018pda} 
with the parameter $\eta=10$, where the ratio 
between the scale-up coordinate interval $\Delta x$ and the Cartesian coordinate interval $\Delta X$  
is given by $\Delta x/\Delta X=1+\eta$ at the center. 
For the mesh refinement, we introduce two upper layers to resolve the gravitational collapse around the center. 
Therefore 4 times finer resolution is realized near the center when they are introduced. 
We start the calculation with the single layer, 
and the upper layers are introduced when the value of the lapse function gets smaller than 0.1 and 0.05 at the origin%
\footnote{\baselineskip5mm
The values at newly introduced grid points are evaluated by the Lagrange interpolation.  
In the upper layer, 7 grid points on the boundary are regarded as buffer points, and 3 of them (every two) are shared with the lower layer. 
In an iteration step of the time evolution, first, the values on the buffer grid points are evaluated by the interpolation 
from the values on the lower layer. 
Then the values at the inner 4 buffer points are evaluated by solving the evolution equations 
while the values at the outer 3 points are evaluated by the interpolation from the values on the lower layer. 
In the inside bulk region, the values obtained by the time evolution are kept 
and taken to the lower layer grid points 
while, for the buffer grid points, 
they are discarded just before the next evolution step of the lower layer. }. 
For the initial condition, we set $a_\ii=1$ and $H_\ii=50/L=5k$. 
The number of grids for each side is taken as 60, 70 and 80 for the initial lowest layer, 
and the convergence is checked. 

\begin{figure}[htbp]
  \begin{center}
  \includegraphics[scale=1.25]{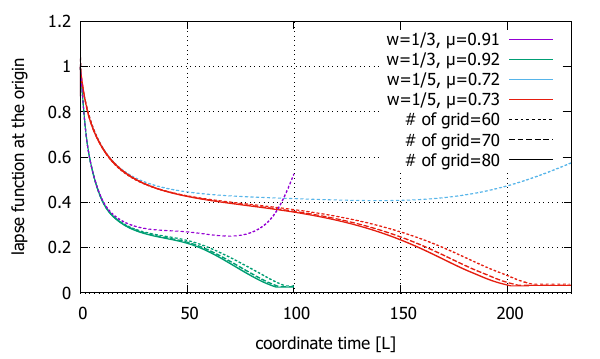}
  \caption{
  \baselineskip5mm  
  The time evolution of the lapse function at the origin. 
  The detection of the horizon triggers the excision procedure inside the horizon and 
  the lapse function at the origin is fixed after that. 
  }
  \label{fig:alpha}
  \end{center}
  \end{figure}

First, let us show the existence of the threshold value of the amplitude $\mu$ for the black hole formation. 
In order to explicitly show the existence of the threshold value, 
we show the time evolution of the value of the lapse function at the origin. 
In our gauge condition for the numerical simulation, the value of the lapse function significantly decreases and approaches zero before the horizon formation. 
On the other hand, if the value of the amplitude $\mu$ is not sufficiently large, 
the value of the lapse function at the origin bounces back and no horizon formation is observed. 
In Fig.~\ref{fig:alpha}, we show the value of the lapse function at the origin as a function of the coordinate time. 
From the behavior of the lapse function, we can see that there are threshold values of $\mu$ between 
0.91 and 0.92 for $w=1/3$ and between 0.72 and 0.73 for the $w=1/5$ case. 

\begin{figure}[htbp]
  \begin{center}
  \includegraphics[scale=1.25]{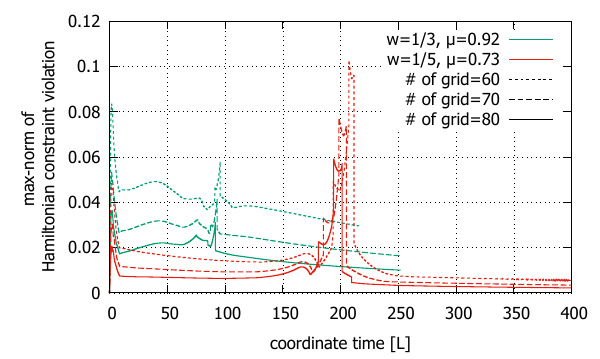}
  \caption{\baselineskip5mm
  Time evolution of the normalized max-norm of the Hamiltonian constraint violation. 
  }
  \label{fig:constraint}
  \end{center}
  \end{figure}

Hereafter we show the results for the case I:$(w,\mu)=(1/3,0.92)$ and 
case II:$(w,\mu)=(1/5,0.73)$, that is, 
the cases slightly above the threshold values. 
First, let us show the violation of the Hamiltonian constraint in Fig.~\ref{fig:constraint}. 
The violation of the Hamiltonian constraint is evaluated at each grid point 
with an appropriate normalization, and the maximum value $H_{\rm max}$ is taken including the grid points on the higher layers if they exist. 
We excluded the grid points well inside the horizon for the evaluation of the constraint violation. 
A discontinuous change can happen when a higher layer is introduced or 
the horizon is formed. 
In addition, depending on how to exclude the grid points inside the horizon, 
the value of $H_{\rm max}$ slightly changes. 
Due to this dependence, a discontinuous change also may be observed when the calculation is terminated once and continued with the data at the termination (it can be seen at around $t=210L$ for the red solid line in Fig.~\ref{fig:constraint}). 
Overall, the constraint violation is acceptably small outside the horizon and 
reasonable convergent behaviors with different resolutions can be found.

\begin{figure}[htbp]
  \begin{center}
  \includegraphics[scale=0.55]{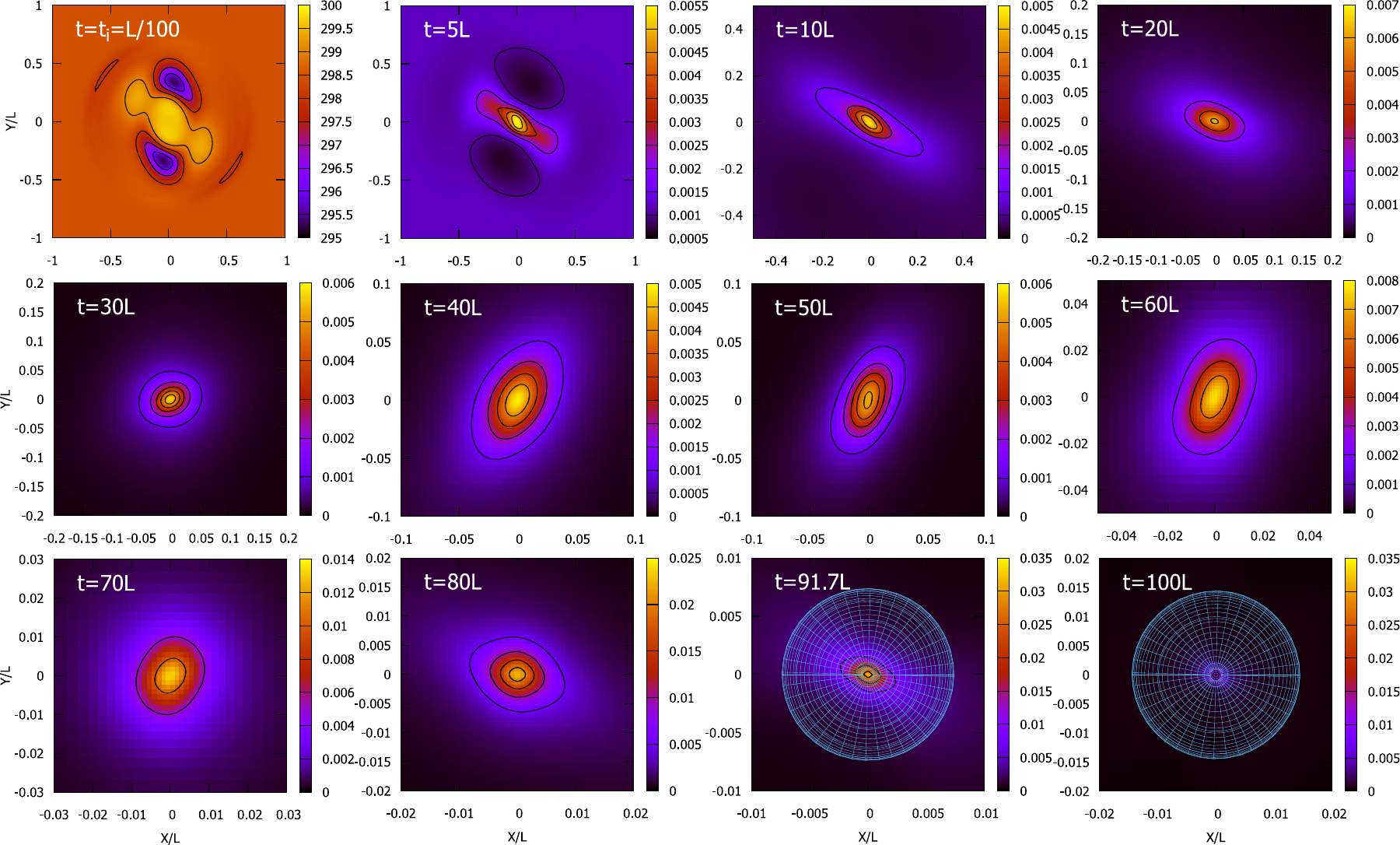}
  \caption{\baselineskip5mm
  Snapshots of the contour map of the fluid comoving density on the $z=Z=0$ plane for the case I:$(w,\mu)=(1/3,0.92)$. The figures are described in the Cartesian coordinates $X$ and $Y$. The green meshes in the last two panels describe the apparent horizon. 
  The horizon is first found at the time $t=91.7L$. 
  }
  \label{fig:coden_rad}
  \end{center}
  \end{figure}

\begin{figure}[htbp]
  \begin{center}
  \includegraphics[scale=0.55]{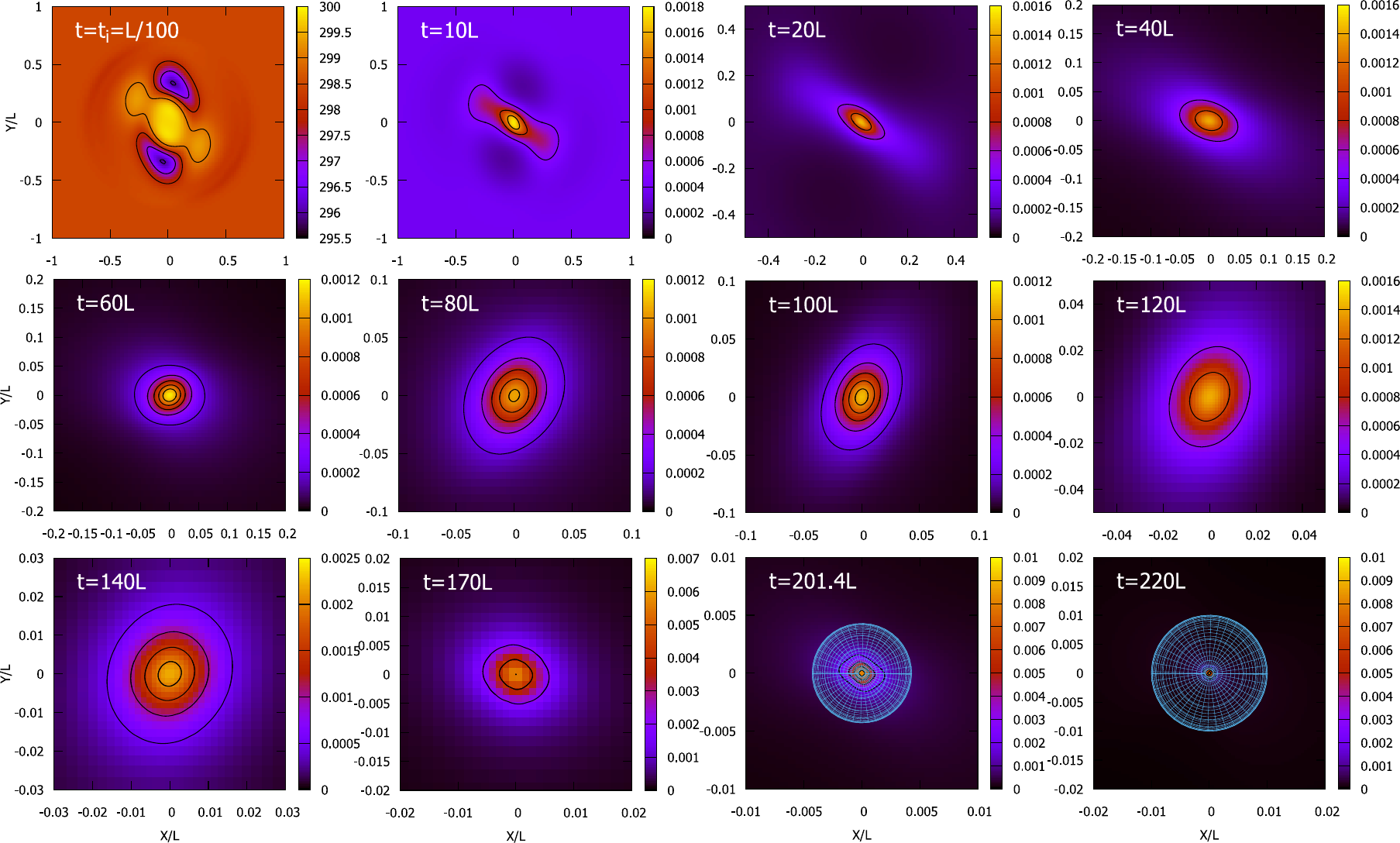}
  \caption{
    \baselineskip5mm
    Snapshots of the contour map of the fluid comoving density on the $z=Z=0$ plane for the case II:$(w,\mu)=(1/5,0.73)$. The figures are described in the Cartesian coordinates $X$ ad $Y$. The green meshes in the last two panels describe the apparent horizon. 
  The horizon is first found at the time $t=201.4L$. 
  }
  \label{fig:coden_w02}
  \end{center}
  \end{figure}

Let us show the snapshots of the contour map of the fluid comoving density on the $z=Z=0$ plane in Figs.~\ref{fig:coden_rad} and \ref{fig:coden_w02}. 
It can be seen that the system is highly non-spherical until the apparent horizon forms. 
The time evolutions of the contour map show oscillating behaviors rather than rotation. 
The two cases (I and II) 
are qualitatively similar to each other if the time scale is appropriately tuned. 

\section{evaluation of spin and asphericity of the horizon}
\label{sec:spin}
In this section, we try to estimate the spin and asphericity of the 
black hole by using geometrical quantities of the horizon. 
In a realistic situation, if we can perform the numerical 
simulation for a sufficiently long time. 
The black hole would approach a stationary asymptotically flat black hole, 
namely, a Kerr black hole,  
because the size of the cosmological horizon becomes much larger than 
the black hole horizon scale and the effect of accretion becomes negligible. 
First, let us check the accretion effect by plotting the time evolution of the 
horizon area $A$. 
We plot the value $\sqrt{A/(4\pi)}/(2M_H)$ rather than $A$ itself noting that $\sqrt{A/(4\pi)}$ has the mass dimension, 
where $M_H$ is the horizon mass defined in Eq.~\eqref{eq:Mh}. 
\begin{figure}[htbp]
  \begin{center}
  \includegraphics[scale=1.25]{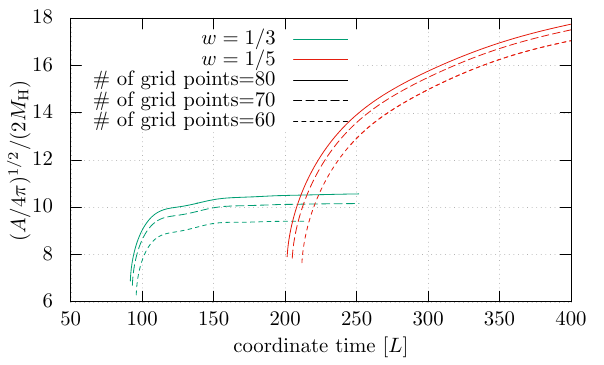}
  \caption{\baselineskip5mm
  Time evolution of the value of $\sqrt{A/(4\pi)}/(2M_H)$, 
  where $M_H$ is the horizon mass defined in Eq.~\eqref{eq:Mh}. 
  }
  \label{fig:sqarea}
  \end{center}
  \end{figure}
For case I, the mass accretion almost ends around $t=150L$. 
For case II, although the area of the horizon is gradually increasing even around $t=400L$, the contribution of the accreting mass is expected to be subdominant for the black hole. 

In order to evaluate the spin of the black hole, 
let us refer to the Kerr black hole with the mass $M$ and the spin parameter $a$. 
The area of the horizon $A_{\rm Kerr}$ is given by 
\begin{equation}
  A_{\rm Kerr}=8\pi (M^2+\sqrt{M^4-a^2M^2}). 
  \label{eq:Akerr}
\end{equation}
The equatorial circumference $d_{\rm Kerr}$ 
and meridional circumference $\ell_{\rm Kerr}$ 
of the Kerr black hole can be calculated as 
\begin{eqnarray}
d_{\rm Kerr}&=&4\pi M, 
\label{eq:ekerr}\\
\ell_{\rm Kerr}&=&4\sqrt{2Mr_+}E\left(\frac{a^2}{2Mr_+}\right), 
\label{eq:lkerr}
\end{eqnarray}
where $r_+=M+\sqrt{M^2-a^2}$ and $E()$ is the complete elliptic integral of the 2nd kind. 
Combining Eq.~\eqref{eq:Akerr} and \eqref{eq:ekerr}, 
we find 
\begin{equation}
  \frac{a^2}{M^2}=\frac{4\pi A_{\rm Kerr}\left(d_{\rm Kerr}^2-\pi A_{\rm Kerr}\right)}{d_{\rm Kerr}^4}. 
\end{equation}
Therefore, by using the horizon area $A$ and the equatorial circumference $d$, 
we define the effective dimensionless spin parameter $s$ as 
\begin{equation}
  s^2:=\frac{4\pi A\left(d^2-\pi A\right)}{d^4}.
\end{equation}
We also define the following indicators for the asphericity of the horizon: 
\begin{eqnarray}
  \alpha&:=&\frac{\ell_{x=0}}{d}, \\
  \beta&:=&\frac{\ell_{y=0}}{d}, 
\end{eqnarray}
where $\ell_{X=0}$ and $\ell_{Y=0}$ are the meridional circumference 
measured on the $X=0$ and $Y=0$ planes, respectvely. 
The deviation of $\alpha$ or $\beta$ from 1 indicates that the horizon is 
non-spherical. 

\begin{figure}[htbp]
  \begin{center}
  \includegraphics[scale=1.25]{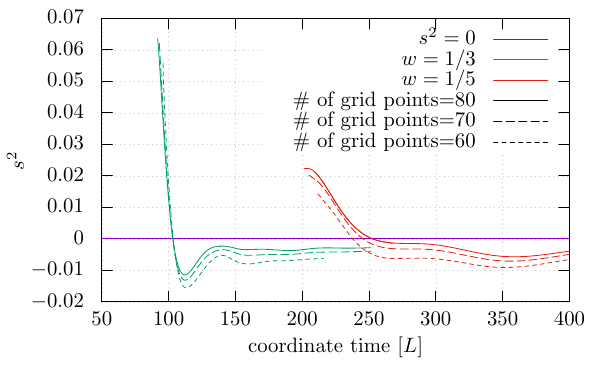}
  \caption{\baselineskip5mm
  Time evolution of the effective dimensionless spin $s^2$. 
  }
  \label{fig:spin}
  \end{center}
  \end{figure}

In Fig.~\ref{fig:spin}, we plot the effective dimensionless spin parameter $s^2$ as a function of the time $t$ for each parameter set. 
It can be found that the value of $s^2$ seems to converge to a negatively small value at a late time. 
The absolute value of $s^2$ tends to be a smaller value for a finer resolution and seems to converge to a very small value $|s^2|\lesssim 10^{-3}$ 
for a sufficiently fine resolution. 
Therefore we conclude that our result is consistent with $s^2=0$ within the numerical precision. 

\begin{figure}[htbp]
  \begin{center}
  \includegraphics[scale=1.25]{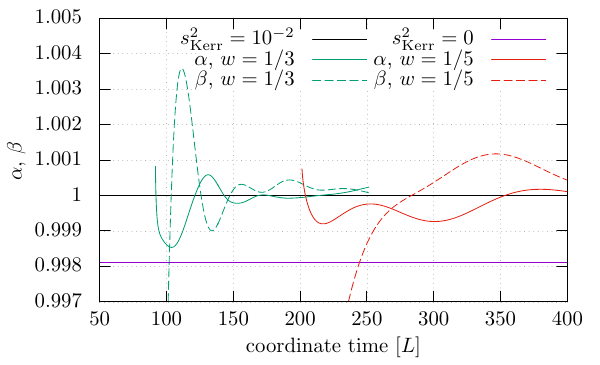}
  \caption{
    \baselineskip5mm
    Time evolution of the values of $\alpha$ and $\beta$ for each parameter set. 
    The solid black line shows the value of $\ell_{\rm Kerr}/d_{\rm Kerr}$ 
    for the case $s_{\rm Kerr}^2:=a^2/M^2=10^{-2}$. 
    The results for the highest resolution (80 grid points for each side in the lowest layer) 
    are used for this figure. 
  }
  \label{fig:circ}
  \end{center}
  \end{figure}

Since our setting is not exactly vacuum and not asymptotically flat, 
the value of $s^2$, which is defined in analogy with the Kerr black hole solution, may not be perfectly suitable for our purpose. 
Thus we also check the values of $\alpha$ and $\beta$, which are the indicators of the asphericity of the horizon. 
If the horizon is highly spherically symmetric, we may not expect the 
large value of the spin parameter. 
As is shown in Fig.~\ref{fig:circ}, the horizon is 
highly spherically symmetric at a late time. 
We also show the value of $\ell_{\rm Kerr}/d_{\rm Kerr}$ for the case $s_{\rm Kerr}^2:=a^2/M^2=10^{-2}$ 
in Fig.~\ref{fig:circ}. 
We can see that the dimensionless spin parameter is expected to be much smaller than $10^{-2}$. 

\section{summary and discussions}
\label{sec:summary}
We performed numerical simulations of primordial black hole formation 
from a non-spherical initial density profile 
with a misaligned deformation tensor. 
The equation of states is assumed to be the linear relation between 
the energy density $\rho$ and the pressure $p$, namely, 
$p=w\rho$ ($w=1/3$ or $1/5$). 
We have optimized the initial profile for the generation of PBH spin due to the tidal torque taking the probability estimation based on the peak theory into account. 
That is, the initial setting is finely tuned to make the tidal torque more effective. 
Then the probability of obtaining such a situation is highly suppressed 
based on the peak theory for PBH formation, and the PBH spin is expected to be larger than the typical value. 
Nevertheless, our results are consistent with non-rotating PBH formation, 
that is, the spin parameter of the resultant PBH is 
so small that we cannot detect the non-zero value of the spin within our numerical precision. 
This result is qualitatively consistent with the prediction 
provided by the analyses based on perturbative approaches~\cite{DeLuca:2019buf,Harada:2020pzb,Saito:2023fpt}. 

It should be mentioned that there are several caveats to our results. 
First, we only performed the numerical simulation for a specific initial inhomogeneity. 
Other initial profiles could give different results. 
Due to the limitation of the computer resources, the support of the initial inhomogeneity 
could not be much smaller than the numerical domain to keep a sufficient resolution. 
Therefore, essentially, in our simulation, the system has been characterized 
by only one scale of $1/k$. 
The contribution of multiple scales might promote angular momentum transfer. 

Even if we stick to the initial configuration treated in this paper, 
for a value of the amplitude $\mu$ very close to the threshold for the black hole formation, 
we would get a very small black hole associated with 
the critical behavior\cite{Choptuik:1992jv,Koike:1995jm}. 
Since a larger spin parameter has been predicted in this critical scaling regime 
from the perturbative analyses\cite{DeLuca:2019buf,Harada:2020pzb,Saito:2023fpt}, 
we may expect to detect a non-zero spin in the case near the threshold.

Our results suggest that, for $p/\rho=w\gtrsim 0.2$, the dimensionless PBH spin $s$ 
is typically so small that $s\ll0.1$. Since the softening due to the QCD crossover 
is expected to satisfy $w>0.2$, the crossover seems not to contribute much to spinning up PBHs. 
One interesting possibility is considering a much softer equation of states $w\ll 0.2$ 
with a beyond-standard model in the universe earlier than electroweak phase transition~\cite{Escriva:2022yaf,Escriva:2023nzn}. 
PBH formation from scalar condensate objects is also an interesting possibility for having highly spinning 
PBHs (see, e.g., Refs.~\cite{Cotner:2018vug,Cotner:2019ykd}). 
For the PBH formation in a matter-dominated universe, 
since the pressure gradient force is absent, asphericity is essential 
in the consideration of the PBH formation criterion~\cite{Harada:2017fjm,Harada:2020pzb}. 
Then we can expect highly spinning PBH formation~\cite{Harada:2020pzb}. 
The simulations of the spinning PBH formation in the early universe with 
the extremely soft equation of states and an early matter-dominated universe 
would be interesting future works. 

\section*{Acknowledgements}
We would like to thank T. Harada for useful comments on the draft. 
This work was supported by JSPS KAKENHI Grant
Numbers JP20H05850 and JP20H05853.

\appendix

\section{Horizon entry time and horizon mass}
\label{sec:horizon_scale}

We calculate the horizon entry time for the inhomogeneity 
whose length scale is given by $1/k$ in the universe filled with 
the perfect fluid with the equation of states given by $p=w\rho$. 
The scale factor and the Hubble parameter behave as functions of the cosmological time $t$ as 
\begin{eqnarray}
  a&=&a_\ii\left(\frac{t}{t_\ii}\right)^{\frac{2}{3(1+w)}}, \\
  H&=&\frac{2}{3(1+w)}\frac{1}{t}, 
\end{eqnarray}
where $t_\ii$ and $a_\ii$ are the initial time and the scale factor, respectively.  
The value of $aH$ is given by 
\begin{equation}
  aH=a_\ii H_\ii\left(\frac{3}{2}(1+w)H_\ii t\right)^{-\frac{1+3w}{3(1+w)}}, 
\end{equation}
where $H_\ii$ is the initial Hubble parameter. 
Then the horizon entry condition $a_{\rm ent}H_{\rm ent}=k$ can be 
rewritten as 
\begin{equation}
  a_{\rm ent}H_{\rm ent}=a_\ii H_\ii\left(\frac{3}{2}(1+w)H_\ii t_{\rm ent}\right)^{-\frac{1+3w}{3(1+w)}}=k.  
\end{equation}
The horizon entry time $t_{\rm ent}$ is given by
\begin{equation}
  t_{\rm ent}=\left(\frac{k}{a_\ii H_\ii}\right)^{-\frac{3(1+w)}{1+3w}}
  \frac{2}{3(1+w)H_\ii}. 
\end{equation}
The horizon mass $M_{\rm H}$ 
can be estimated as follows:
\begin{eqnarray}
  M_{\rm H}=\frac{1}{2}H^{-1}_{\rm ent}&=&\frac{3}{4}(1+w)t_{\rm ent}
  =\frac{1}{2H_\ii}\left(\frac{k}{a_\ii H_\ii}\right)^{-\frac{3(1+w)}{1+3w}}. 
  \label{eq:Mh}
\end{eqnarray}


\end{document}